# Universal High Order Subroutine with New Shock Detector for Shock Boundary Layer Interaction

M. Oliveria[1], P. Lu[2], X. Liu[3] and C. Liu[4]
*University of Texas at Arlington, Arlington, TX 76019*

The goal of this work is to develop a new universal high order subroutine for shock boundary layer interaction. First, an effective shock/discontinuity detector has been developed. The detector has two steps. The first step is to check the ratio of the truncation errors on the coarse and fine grids and the second step is to check the local ratio of the left and right slopes. The currently popular shock/discontinuity detectors such as Harten's, Jameson's and WENO can detect shock, but mistake high frequency waves and critical points as shock and then damp the physically important high frequency waves. Preliminary results show the new shock/discontinuity detector is very delicate and can detect all shocks including strong, weak and oblique shocks or discontinuity in function and the first, second, and third order derivatives without artificial constants, but never mistake high frequency waves and critical points, expansion waves as shock. This will overcome the bottle neck problem with numerical simulation for the shock-boundary layer interaction, shock-acoustic interaction, image process, porous media flow, multiple phase flow and anywhere the high frequency waves are important, but discontinuity exists and is mixed with high frequency waves. After detecting the shock we can then use one side high order scheme for shocks and high order central compact scheme for the rest if the shock is appropriately located. Then a high order universal subroutine for finite difference method is developed which can be used for any finite difference code for accurate numerical derivatives.

## Nomenclature

| | | |
|---|---|---|
| $E, F, G$ | = | Fluxes in Nevier-Stokes Equations |
| $\xi, \eta, \zeta$ | = | Curvilinear Coordinates |
| $IS$ | = | WENO smoothness |
| $f$ | = | Input function |
| $\rho, u, P$ | = | Density, velocity, and pressure |
| $T$ | = | Truncation Error |
| $t$ | = | Time |
| $i$ | = | Index in x-direction |

## I. Introduction

**1.1 Importance of Shock/discontinuity detector**

The high order compact finite difference scheme (Lele 1992) has been widely applied for numerical simulation of flow transition, turbulence, acoustic problems, and many other places where the high order accuracy and high resolution is important. However, the high order compact scheme will meet severe trouble when the shock/discontinuity occurs in the computational domain since the continuity in function and/or derivatives are lost and the classical numerical analysis, which is based on an assumption that the subject function is continuous in function and any order of derivatives, will all fail. Therefore, an effective shock/discontinuity detector which can detect any shock including strong shock, weak shock, oblique shock or discontinuity in function, first order derivative, second order derivative, etc., is critical to the success of numerical simulation for shock-turbulence interaction, shock-acoustic interaction, porous media flow, multiple phase flow, image process, etc., where both shock and high frequency waves are important. Recently, many different numerical methods have been used to accomplish this, with combinations of high order numerical schemes and filters (Martin, 2007; Visbal & Gaitonde, 2002; Jiang et al, 2001). The application of low-order filters on areas with shock waves to remove non-physical wiggles can achieve success only when the shock locator can give the right locations for shock/discontinuity. Therefore, the development of an effective shock locator is critical.

---

[1] PhD student, Department of Mathematics, Box 19408, UTA, Arlington, TX 76019.
[2] PhD student, Department of Mathematics, Box 19408, UTA, Arlington, TX 76019.
[3] PhD student, Department of Mathematics, Box 19408, UTA, Arlington, TX 76019.
[4] Professor, Department of Mathematics, Box 19408, UTA, Arlington, TX 76019, AIAA Associate Fellow.

The current popular shock/discontinuity detectors including Harten's (1978), Jameson's (1981), WENO (1996) use first derivatives and/or second order derivatives of the function to detect shock. They can detector shock/discontinuity effectively, but mistake high frequency sinuous functions and critical point as shock since the first derivative is high for high frequency waves and the second order derivative is high for critical points. How to distinguish shock/discontinuity from high frequency waves and critical points was though extremely difficult if not impossible. There are many new efforts (Zhou et al, 2007; Cueto-Felguerroso et al 2007; Archibald et al, 2008) on this challenging topic.

Unlike other shock locators, this effort is to develop a very efficient shock detector which can detect any strong or weak discontinuity in function, first order, second order, and third order derivatives, but never mistakes high frequency sinuous functions as shock at reasonable cost in computation. With preliminary successes for 9 test cases for the high frequency waves, mixed high frequency wave with discontinuity, large slope smooth function, shock tube, and shock-entropy interactions, the new two step shock detector has not found any particular case which can make the new detector fail.

This new shock detector is a two step check with global multilevel grids truncation error ratio check and local left and right slope ratio check (details can be found in the following sections).

After we detect shock, we can use high order central compact scheme for the smooth part, but one side high order scheme for the shock/discontinuity or low order filter for the shock area to eliminate adjunct wiggles around the shock.

**1.2 Fundamental tasks for finite difference methods**

The fundamental task of finite difference methods for ODE or PDE is to provide an accurate approximation of derivatives for a given discrete data set. Traditional finite difference schemes such as central or up-winding schemes are developed based on analysis of physics, for example, central for elliptic and up-winding for hyperbolic problems. On the other hand, the computer does not know physics. It only accepts a set of discrete data as an input and gives back a set of discrete data as output derivatives, without consideration of the problems potentially stemming from fluid dynamics, biology or social science. Therefore, our efforts are focused on how to provide an accurate derivative for a given data set. To do this, we have to first detect the shock/discontinuity and then use the one side high order scheme for shock and high order central scheme for the smooth part.

**1.3 Importance of high order scheme to DNS/LES**

It should be pointed out that the order of accuracy of a finite difference scheme is absolutely not a trivial issue to CFD, especially to direct numerical simulation (DNS) and large eddy simulation (LES). There is a big difference in requirements of grid size by DNS/LES between low order schemes and high order schemes. Let us take a look at the local truncation error for 1-D problem. If one uses a second order scheme with a mesh size of $\Delta x_2$ and wants to have the same truncation error as a sixth order scheme with a mesh size of $\Delta x_6$, one should have:

$$C_2(\Delta x_2)^2 = C_6(\Delta x_6)^6. \tag{1.1}$$

Assume $C_2 \approx C_6$ and $\Delta x_6 = 0.01$ (100 grid points in a normalized domain), we will get $(\Delta x_2)^2 = (10^{-2})^6$,

$$\Delta x_2 = 10^{-6}. \tag{1.2}$$

In other words, the second order scheme needs one million of grid points to beat the sixth order scheme with 100 grid points for the same accuracy. This advantage of high order scheme will become more significant when one uses

DNS for 3-D problems. We cannot use, and do not want to use, millions of grids in each direction for DNS, but prefer to use 100 grid points. Therefore the high order scheme must be used. Of course, the global error does not only depend on the local truncation error and $C_2 \neq C_6$. The advantage of the sixth order scheme does not show 10 thousand times better than the second order scheme. However, it is now widely recognized that high order finite difference schemes are strongly encouraged to be used for DNS and LES which have much higher accuracy and higher resolution with the same grid size than low order schemes do.

### 1.4 New point of view on high order CFD

The 3-D time dependent Navier-Stokes equations in a general curvilinear coordinate can be written as

$$\frac{1}{J}\frac{\partial Q}{\partial t} + \frac{\partial(E - E_v)}{\partial \xi} + \frac{\partial(F - F_v)}{\partial \eta} + \frac{\partial(G - G_v)}{\partial \zeta} = 0 \quad (1.3)$$

For 1-D conservation law, it will be:

$$\frac{\partial Q}{\partial t} + \frac{\partial E}{\partial \xi} = 0 \quad (1.4)$$

The critical issue for high order CFD is to find an accurate approximation of derivatives for a given discrete data set. The computer does not know any physical process but rather a discrete data set. The output derivative is also a discrete data set. For three dimensional problems, we only need to call the universal subroutine three times.

### 1.5 A short overview on shock capturing schemes

The flow field is in general governed by the Navier-Stokes system. In the smooth region of solutions, the Navier-Stokes system presents parabolic type behavior and is therefore dominated by viscosity or second order derivatives. One expects that the equation should be solved by the high order central difference scheme, high order compact scheme is preferable, to get high order accuracy and high resolution. High order of accuracy is critical in resolving small length scales in the flow transition and turbulence process. However, for external flow, the viscosity is important largely only in the boundary layers. The main flow can still be considered as inviscid and the governing system can be dominated by the time dependent Euler equations which are hyperbolic. The difficult problem with the numerical solution is shock capturing which can be considered as a discontinuity or mathematical singularity (no classical unique solution and no bounded derivatives). In the shock area, continuity and differentiability of the governing Euler equations are lost and only the weak solution in an integration form can be obtained. The shock can be developed because the Euler equation is non-linear and hyperbolic. For the hyperbolic system, the analysis already shows the existence of characteristic lines and Riemann invariants. Apparently, the upwind finite difference scheme coincides with the physics for a hyperbolic system. History has shown the great success of upwind technologies. Many upwind or bias upwind schemes have achieved great success in capturing the shocks sharply, such as Godnov (1959), Roe (1981), MUSCL (Van Leer, 1979), TVD (Harten, 1983), ENO (Harten et al, 1987; Shu et al, 1988, 1989) and WENO (Liu et al, 1994; Jiang et al, 1996). However, all these shock-capturing schemes are based on upwind or bias upwind technology, which is nice for a hyperbolic system, but is not favorable to the N-S system which presents parabolic equation behavior. The small length scale is very important in the flow transition and turbulence process and thus very sensitive to any artificial numerical dissipation. High order compact scheme (Lele, 1992; Visble, 2002) is more appropriate for simulation of flow transition and turbulence because it is central and non-dissipative with high order accuracy and high resolution. Due to the usage of derivatives, compact schemes usually give us a tri-diagonal or penta-diagonal system. Although the tri-diagonal matrix is sparse, the inverse of a tri-diagonal matrix is dense, which means the derivative at a certain grid point depends upon all the grid points along a grid line. The success of compact schemes indicates that the global dependency is very important for high resolution. However, the global dependency is good for resolution but not so applicable for shock capturing.

The basic idea proposed in ENO (Harten et al, 1987) and WENO (Jiang et al, 1996) schemes is to avoid the stencil containing a shock. ENO chooses the smoothest stencil from several candidates to calculate the derivatives. WENO controls the contributions of different stencils according to their smoothness. In this way, the derivative at a certain grid point, especially one near the shock, is dependent on a very limited number of grid points. The local dependency here is favorable for shock capturing and helps to obtain the non-oscillatory property. The success of ENO and WENO schemes indicates that the local dependency is critical for shock capturing.

Apparently, the critical point is to find the exact location of the shock/discontinuity, or, in other words, to develop a very effective shock/discontinuity locator. After we find the location of the shock, we can then use one-side finite differences for the shock point and high order central finite differences for the smooth part or use the second order filter to eliminate the wiggles around the shock

## II. New Shock/Discontinuity Detector

To introduce our new two step shock/discontinuity detector, we need to introduce some popular shock detectors first.

1) **Harten's Switch Function and Jameson's Shock Detector**

   Harten (1978) defined an automatic switch function that is able to detect large changes in the variation of the function values $f_i$. It generates values between 0 and 1, where 0 is considered smooth and 1 is considered non-smooth.

   The switch is defined as
   $$\theta_{j+1/2} = \max(\hat{\theta}_j, \hat{\theta}_{j+1}), \tag{2.1}$$
   with
   $$\hat{\theta}_i = \begin{cases} \left\| \dfrac{|\alpha_{i+1/2}| - |\alpha_{i-1/2}|}{|\alpha_{i+1/2}| + |\alpha_{i-1/2}|} \right\|^p, & \text{if } |\alpha_{i+1/2}| + |\alpha_{i-1/2}| > \varepsilon \\ 0, & \text{otherwise} \end{cases}$$

   where $\alpha_{i+1/2} = f_{i+1} - f_i$ and $\varepsilon$ is a suitably chosen measure of insignificant variation in $f$.

   Jameson's (1981) shock detector is similar, which can be described as:

   $$v_i = \frac{|p_{i-1} - 2p_i + p_{i+1}|}{|p_{i-1}| + 2|p_i| + |p_{i+1}|} \tag{2.2}$$

   which is related to the second order derivative of the pressure.

2) **WENO**

   The WENO weights use smoothness measurements that evaluate the changes in the variation of the function values $f_i$. Assuming that the three weights have equal contribution, we can determine that a function is smooth if all values are approximately 1/3.

   $$\omega_i = \frac{\alpha_i}{\sum_j \alpha_j}; \quad \alpha_i = \frac{1}{(IS_i + \varepsilon)^2}, \quad i = 0,1,2 \quad \text{where} \tag{2.3}$$

   $$IS_0 = \frac{13}{12}(f_{i-2} - 2f_{i-1} + f_i)^2 + \frac{1}{4}(f_{i-2} - 4f_{i-1} + 3f_i)^2;$$
   $$IS_1 = \frac{13}{12}(f_{i-1} - 2f_i + f_{i+1})^2 + \frac{1}{4}(f_{i-1} - f_{i+1})^2; \tag{2.4}$$
   $$IS_2 = \frac{13}{12}(f_{i+2} - 2f_{i+1} + f_i)^2 + \frac{1}{4}(f_{i+2} - 4f_{i+1} + 3f_i)^2.$$

3) **New Two Step Shock/Discontinuity Locator**

Step 1: Determine the multigrid ratio of the approximation of the sum of the 4$^{th}$, 5$^{th}$ and 6$^{th}$ truncation errors for [$F = f$ + smooth sine wave of small amplitude] and select the points where the ratio is smaller than 4. Theoretically, the ratio of 4$^{th}$ order truncation error of coarse and fine grids should be 16, but any function has a ratio of 4 will be considered smooth and passing the test. The points which have a ratio less than 4 will be picked out for the second left and right slope ratio check.

The multigrid truncation error ratio check is:

$$MR(i,h) = \frac{T_C(i,h)}{T_F(i,h) + \varepsilon},  \qquad (2.5)$$

where $T_F(i,h)$ is the truncation error sum (4$^{th}$, 5$^{th}$, and 6$^{th}$) calculated at the fine grid with $n$ points, $T_C(i,h)$ is the truncation error sum calculated at the coarse grid with n/2 points

Step 2: Calculate the local left and right slope ratio check only at the points which have first ratio less than 4.

The new local left and right slope ratio check is:

$$LR(i) = \frac{\left\| \frac{f'_R}{f'_L} \right| - \left| \frac{f'_L}{f'_R} \right\|}{\left| \frac{f'_R}{f'_L} \right| + \left| \frac{f'_L}{f'_R} \right| + \varepsilon} = \left| \frac{(f'_R)^2 - (f'_L)^2}{(f'_R)^2 + (f'_L)^2 + \varepsilon} \right|$$

$$LR(i) = \frac{|f'_L / f'_R| - |f'_R / f'_L|}{|f'_L / f'_R| + |f_R / f_L|} = \left| \frac{\alpha_R^2 - \alpha_L^2}{\alpha_R^2 + \alpha_L^2 + \varepsilon} \right|, \qquad (2.6)$$

where $f'_R = 3f_i - 4f_{i+1} + f_{i+2}$, $f'_L = 3f_i - 4f_{i-1} + f_{i-2}$ and $\varepsilon$ is a small number to avoid division by zero.

Optional step three: Use a cutoff value of 0.8 to create a 0/1 switch function on the result of Step 2. If the value is zero, $f$ is considered locally smooth, and if the value is one, $f$ is shock/discontinuity at that point.

Note that Liu's first step always checks $f + \sigma \sin(k\pi x + \phi)$ instead of f, where $\sigma$ is a small number. Since the all derivatives are calculated by a subroutine with standard compact scheme, the cost of two step checks is relatively inexpensive.

In order to find universal formula, we need to normalize the data set, u(i), i=1, …, n:

$$u_{diff} = | u_{max} - u_{min} | \qquad (2.7)$$

$$\bar{u} = (u - u_{min}) / u_{diff} \qquad (2.8)$$

Here, $u_{max}$ and $u_{min}$ are the maximum and minimum values of u respectively and $\bar{u}$ is normalized. For simplicity, we disregard the hat of u and use u (i) as the normalized data set. However, this normalization is for finding the shock locator only not for the function itself which we calculate for derivatives.

### III. Computational Results of the New Shock Detector

Since we claim the new shock detector can detect discontinuity for function, first, second, and third derivatives for any function, we need to select a wide range of functions for test. Here we select nine cases to compare our new shock detector with popular Harten's detector and WENO detector.

1) Jump function: $f(x) = \begin{cases} 0, & -1 \leq x \leq 0 \\ 1, & 0 < x \leq 1 \end{cases}$, $n = 81$ (3.1)

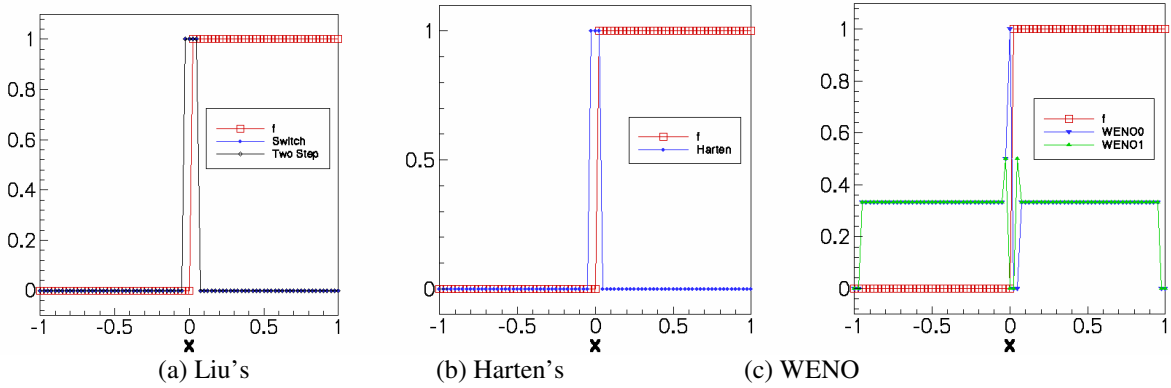

(a) Liu's      (b) Harten's      (c) WENO

Figure 1. Shock detector for jump function

For jump function, all three shock detectors work.

2) Jump slope: $f(x) = \begin{cases} 1+x, & -1 \leq x \leq 0 \\ 1, & 0 < x \leq 1 \end{cases}$, $n = 81$ (3.2)

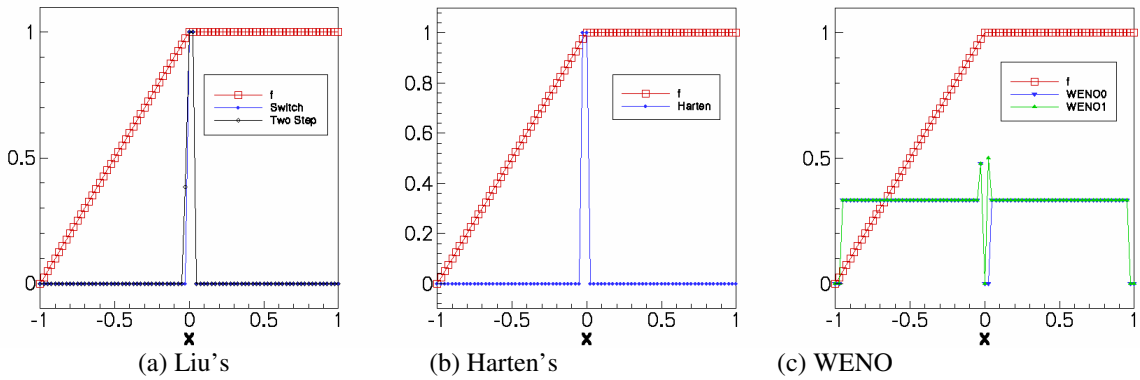

(a) Liu's      (b) Harten's      (c) WENO

Figure 2. Shock detector for jump slope

For jump slope, all three shock ditectors work.

3) High frequency sound waves (eight ponts per wave): $f(x) = \sin\left(\frac{(n-1)\pi x}{8}\right)$, $-1 \leq x \leq 1, n = 81$ (3.3)

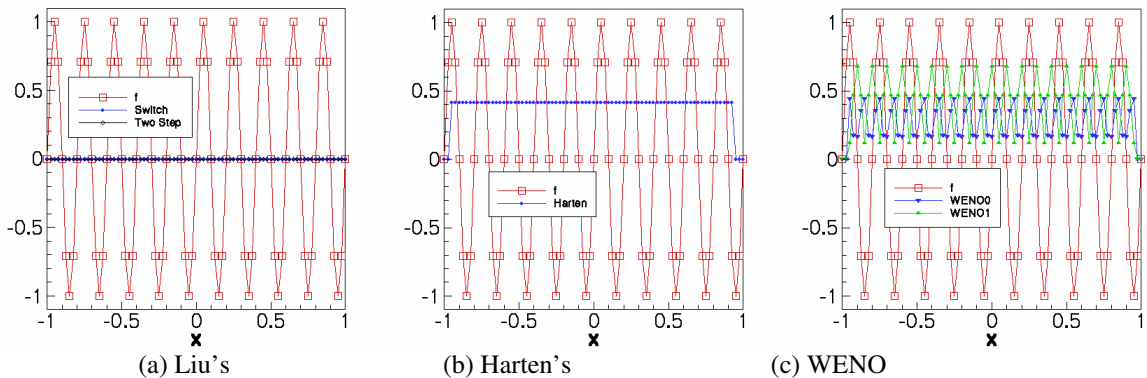

(a) Liu's      (b) Harten's      (c) WENO

Figure 3. Shock detector for high frequncy sound waves

For high frequency sound waves (eight grid points for each wave), both Harton and WENO face serious obstacles and treat the sound waves as shock, but our new detector finds they are all smooth

4) Mixed high frequency sound waves: $f(x) = \sin\left(\frac{(n-1)\pi x}{7}\right) + \sin\left(\frac{(n-1)\pi x}{9}\right)$, $-1 \leq x \leq 1, n = 81$ (3.4)

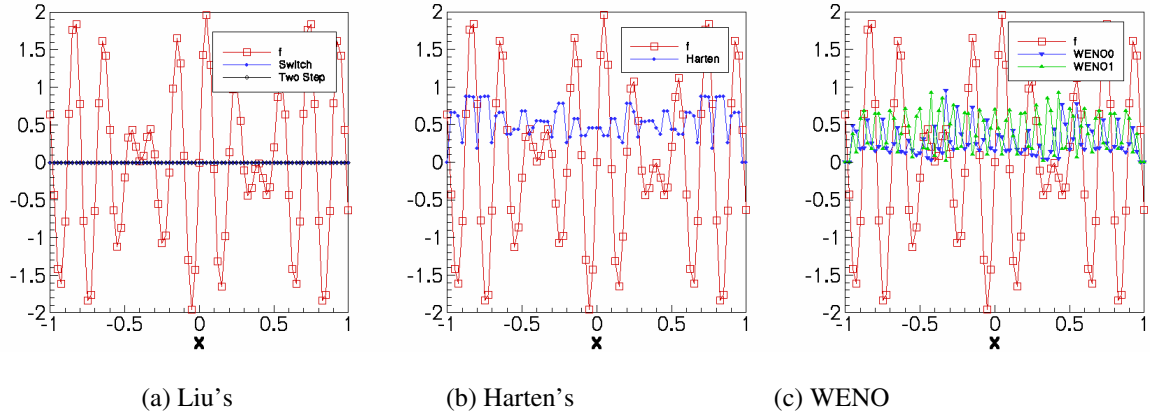

(a) Liu's        (b) Harten's        (c) WENO

Figure 4. Shock detector for mixed high frequncy sound waves

For mixed high frequency waves, Harten's mistreats them as shocks and WENO finds most of them are shocks, wheras but our new detector finds all points are smooth. Note that for WENO the smooth case should have WENO0~WENO1~1/3.

5) Smooth function with large slope: $f(x) = \exp(-300x^2)$, $-1 \leq x \leq 0, n = 81$ (3.5)

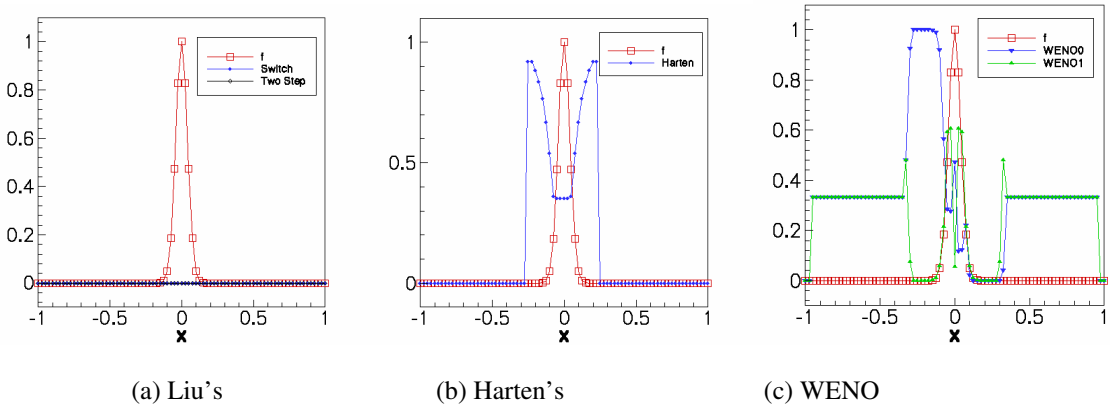

(a) Liu's        (b) Harten's        (c) WENO

Figure 5. Shock detector for smooth function with large slope

For the smooth exponetial function with large slope, both Harten and WENO have serious troubles and mistakenly treat it as a shock, but our new ditector finds it is a smooth function. Harten's and WENO will seriously smear the solution at the bottom. For this case WENO is worse than Harten's.

6) Smooth function with two jumps: $f(x) = \begin{cases} \sqrt{1-\left(\frac{10}{3}x\right)^2}, & -\frac{3}{10} \leq x \leq \frac{3}{10}, n = 81 \\ 0, & \text{otherwise} \end{cases}$ (3.6)

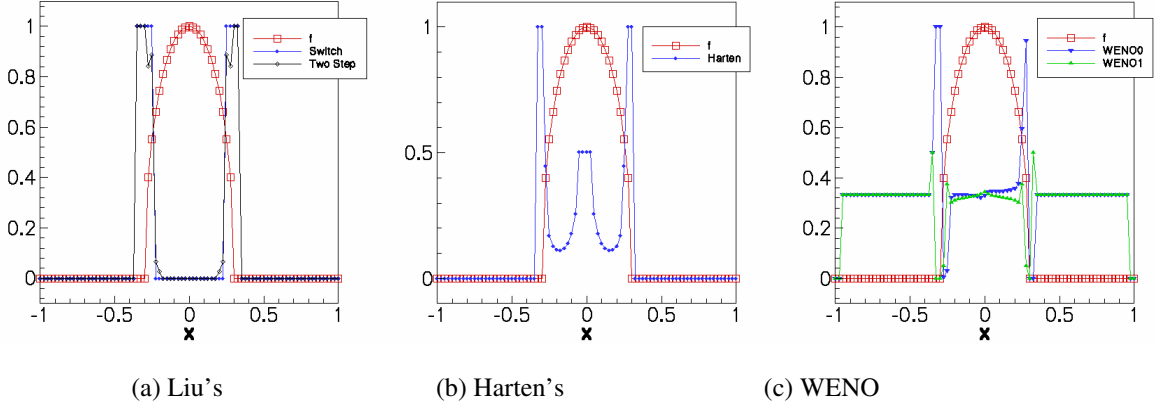

(a) Liu's  (b) Harten's  (c) WENO

Figure 6. Shock detector for smooth function with two jumps

For this mixed function with two jump points at each conner, Harton's mistreats the critical point as shock and WENO captures more points, but ours just finds the exact jump points.

7) Medium frequency with two jump points: $f(x) = \begin{cases} \sin\left(\dfrac{(n-1)\pi x}{16}\right), & -1 \leq x \leq 0 \\ 1 + \sin\left(\dfrac{(n-1)\pi x}{16}\right), & 0 < x \leq 1 \end{cases}$, $n = 81$  (3.7)

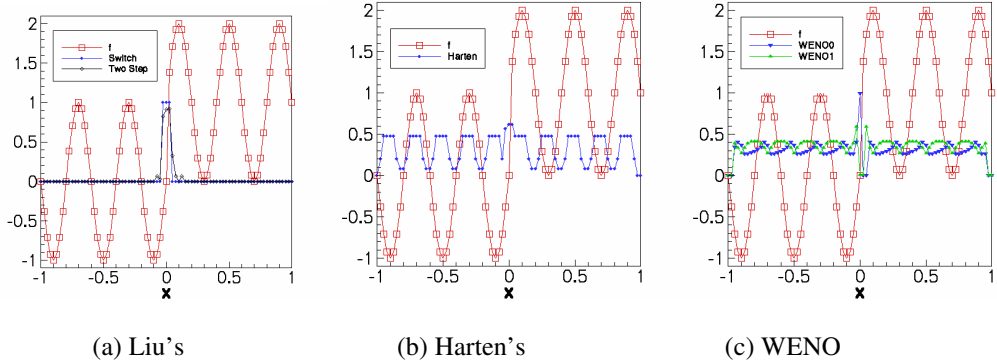

(a) Liu's  (b) Harten's  (c) WENO

Figure 7. Shock detector for smooth function with two jump points

For a medium frequency sound wave with two jump points, Harton's and WENO had trouble distictinguishing the shock and high frequecy waves, but ours finds three points are shock area and the rest are smooth.

8) 1-D Shock Tube Problem (T=2.0, n=101)

In order to use our new detector to detect shocks, we studied the 1-D shock tube case.

The governing equations are 1D Euler equations:

$$\frac{\partial U}{\partial t} + \frac{\partial F}{\partial x} = 0 \text{ where}$$

$$U = (\rho, \rho u, E)^T ; \quad F = (\rho, \rho u + p, u(E + p))^T \tag{3.8}$$

The initial conditions are given as follows:

$$(\rho, u, p) = \begin{cases} (1, 0, 1), & x < 0; \\ (0.125, 0, 0.1) & x \geq 0. \end{cases} \quad (3.9)$$

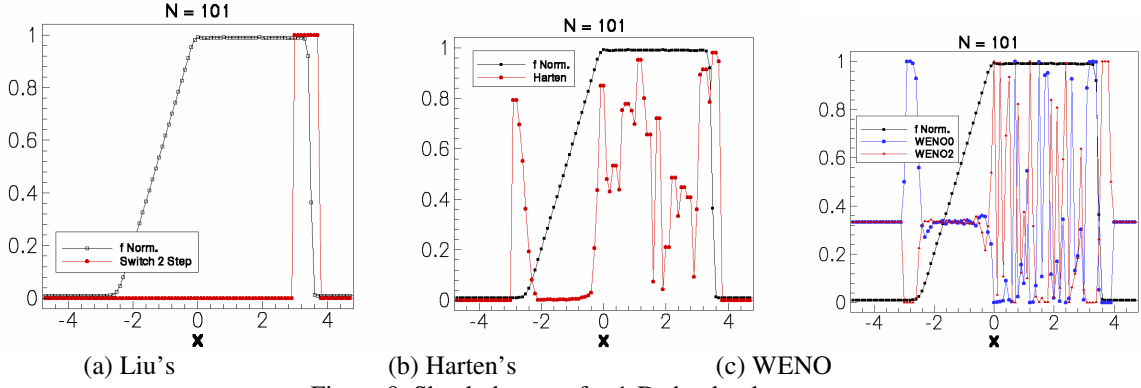

(a) Liu's  (b) Harten's  (c) WENO

Figure 8. Shock detector for 1-D shock tube

All three detectors can capture shock, but Harten's and WENO mistreat the expansion waves as shocks.

9) 1-D Shock/Entropy Wave Interaction (T=1.8, N=201)

To test the capability of the new shock detector, we applied it to the 1-D problem of shock/entropy wave interaction. In this case, 1-D Euler equations are solved with the following initial conditions:

$$(\rho, u, p)_0 = \begin{cases} (3.857143, 2.629369, 10.33333), & x < -4; \\ (1 + 0.2\sin(5x), 0, 1) & x \geq -4. \end{cases} \quad (3.10)$$

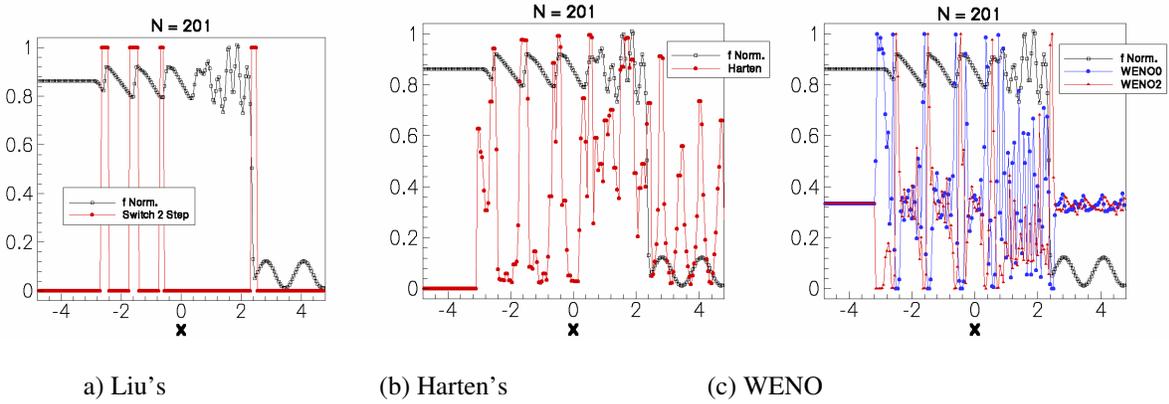

a) Liu's  (b) Harten's  (c) WENO

Figure 9. Shock detector for 1-D shock entropy interaction

All three detectors can capture the shocks including weak shocks, but Harten's and WENO treat the sound waves as shock. That is the reason why WENO smears the sound wave so badly when the grids are not find enough.

## IV. Universal High Order Subroutine

As addressed in the introduction, after reconstruction, the remained problem is to find a more accurate derivative for a discrete data set. Based on this understanding, we developed a black box type subroutine which requires giving an input data set, the data dimension, and finite difference direction. The subroutine will give back a discrete data as an output derivative. The idea is that this subroutine can work for any input data no mater it is smooth, oscillatory, or containing non-differentiable points. Of course, the subroutine should provide high order accuracy when the data is smooth. A black box type subroutine has been developed, applied in our own code, and delivered to Air Force Research Lab. The subroutine is called MWCS(F, FDER, N, Idir) where F is the input function, FDER is the numerical derivative of F, N is the dimension of F, and Idir is for the flux splitting. This black box can be used anywhere or any finite difference CFD code. Of course, this is the first version and need to be improved. For shock and boundary layer interaction, a 3-step TVD Runge-Kutta time marching and a modified $6^{th}$ order weighted compact scheme is applied (Xie et al, 2008)

9American Institute of Aeronautics and Astronautics
092407

### 4.1 Comparison of WENO and MWCS

1) Shock tube
Let first take a shock tube problem to see how the WENO works. The governing equations are 1D Euler equations:

$$\frac{\partial U}{\partial t} + \frac{\partial F}{\partial x} = 0 \qquad (4.1)$$

$$U = (\rho, \rho u, E)^T; \quad F = (\rho u, \rho uu + p, u(E+p))^T$$

The initial conditions are given as follows:

$$(\rho, u, p) = \begin{cases} (1, 0, 1), & x < 0; \\ (0.125, 0, 0.1) & x \geq 0. \end{cases} \qquad (4.2)$$

To solve the Euler equations, Three step TVD Runge-Kutta is used in time marching and Lax-Friedrich flux vector splitting is used and the derivatives of splitting flux $F^+, F^-$ are calculated using WENO. From Figure 10, the shock is smeared before and after shock and the expansion wave is smeared and deformed. This shows WENO has too much dissipation and mistreated the expansion wave. We then use modified WCS and compare with WENO in Figure 11. Apparently, shock is sharper and expansion wave deformation is removed.

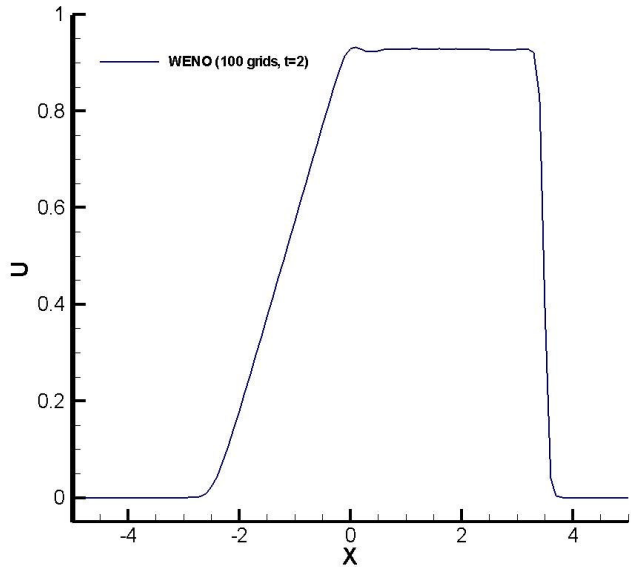

Figure 10 WENO for 1-D shock tube (T=2, grid=100)

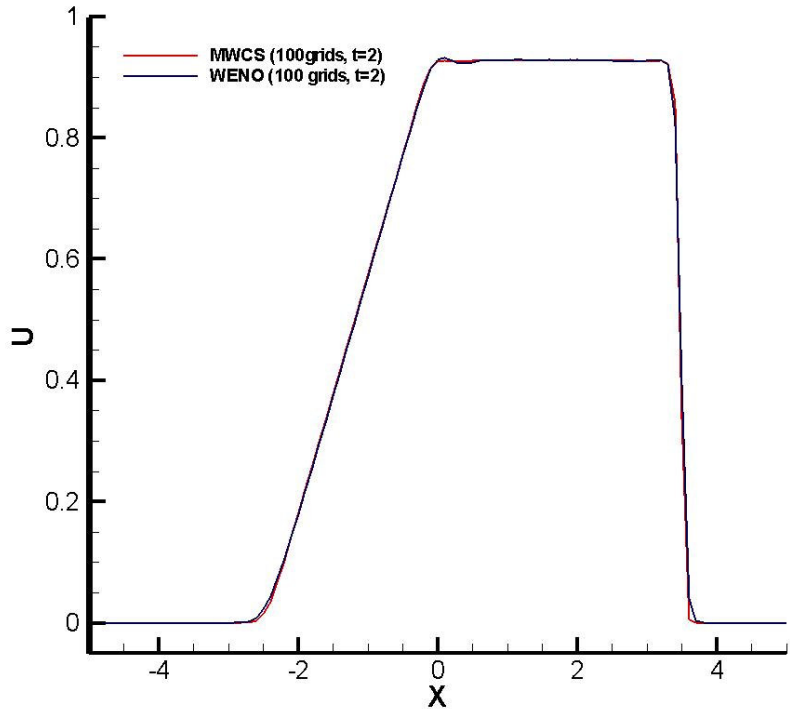

Figure 11 Modified WCS vs WENO for 1-D shock tube (T=2 and grids=100)

2) Shock-entropy interaction

To test the capability of the new scheme in both shock capturing and resolution, we applied it to the 1-D problem of shock/entropy wave interaction. In this case, Euler equations (4.1) are solved with the following initial conditions:

$$(\rho, u, p)_0 = \begin{cases} (3.857143, 2.629369, 10.33333), & x < -4; \\ (1 + 0.2\sin(5x), 0, 1) & x \geq -4. \end{cases} \quad (4.3)$$

On a coarser grid with grid number of N=200, the MWCS (LJX) scheme shows much better resolution for small length scales than the $5^{th}$ order WENO (Figure 12 (a), (b)). This is because MWCS uses central, non-dissipative, compact scheme with weights near the shock area and recovers high order compact right off the shock. The numerical results by our MWCS scheme with 200 grid points are even comparable with the $5^{th}$ order WENO scheme with 1600 grid points (Figure 12 (b)).

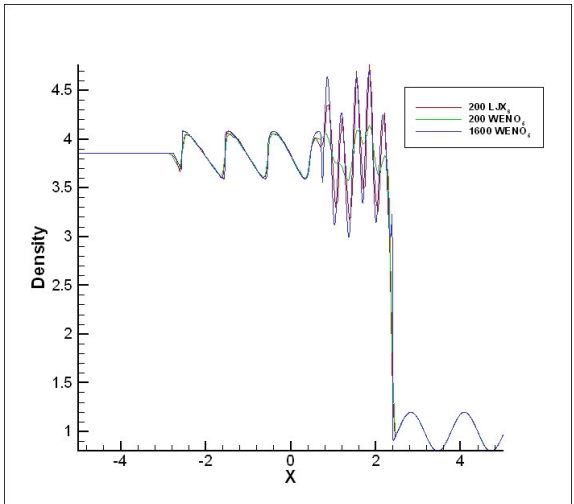
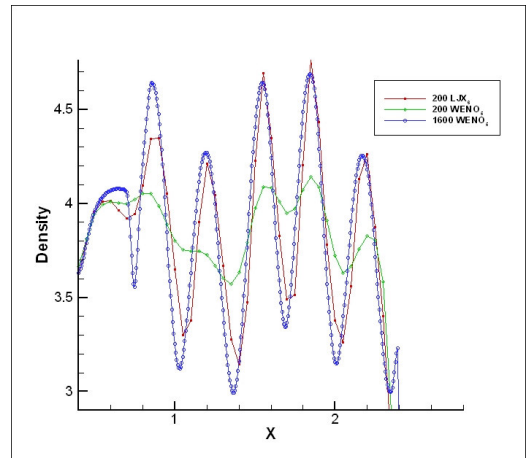

(a)                                                       (b)

Figure 12 Numerical test for 1D shock-entropy wave interaction problem, t=1.8, N=200. (b) is locally enlarged

### 4.2 Preliminary results for 2-D shock-boundary layer interaction

In order to test if current MWCS scheme can work for 2-d and 3-D shock – boundary layer interaction, a 2-D incident shock-boundary layer interaction (Figure 13) was studied by the MWCS. The Reynolds number is $10^5$ and the Mach number was set to 2.15. The overall pressure ratio is 1.55. For comparison, the inflow condition was set as same investigated by Degrez et al (1987). Their experimental work has shown the shock-boundary layer interaction is laminar and two-dimensional. Therefore, we can do a 2-D numerical simulation and compare with their computational and experimental results. The computational grids is 257x257 (Figure 14). The grid stretching in stremwise direction is 1.01. The stretching in normal direction is 1.015. A 2-D Navier-Stokes equation is solved as the governing equation.

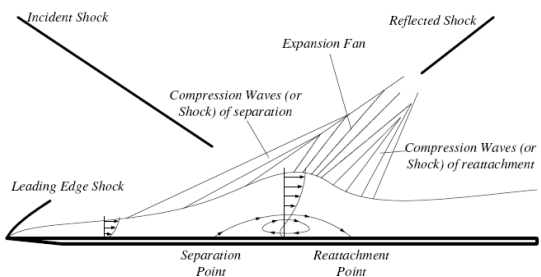
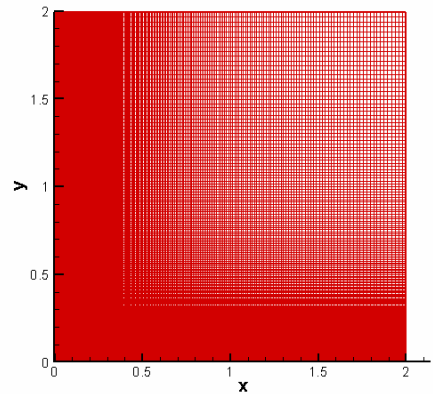

Figure 13 Sketch of incident shock-boundary layer interaction     Figure 14 Computation Grids (257x257)

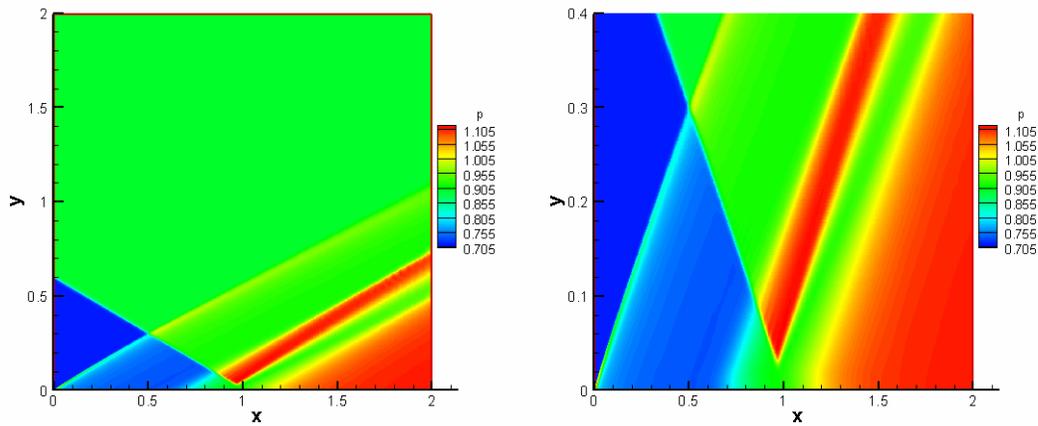

Figure 15  Pressure distribution: left is a normal view and right is stretched in the normal direction by a factor of 5

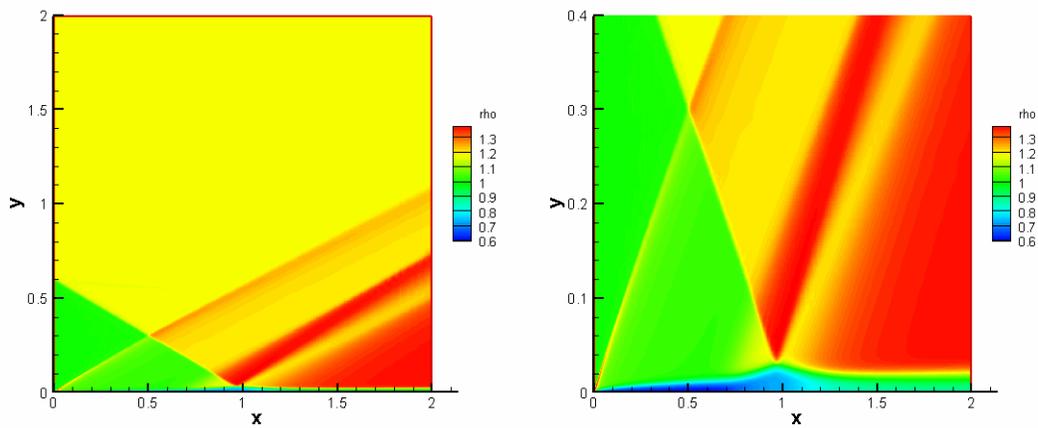

Figure 16  Density distribution: left is a normal view and right is stretched in the normal direction by a factor of 5

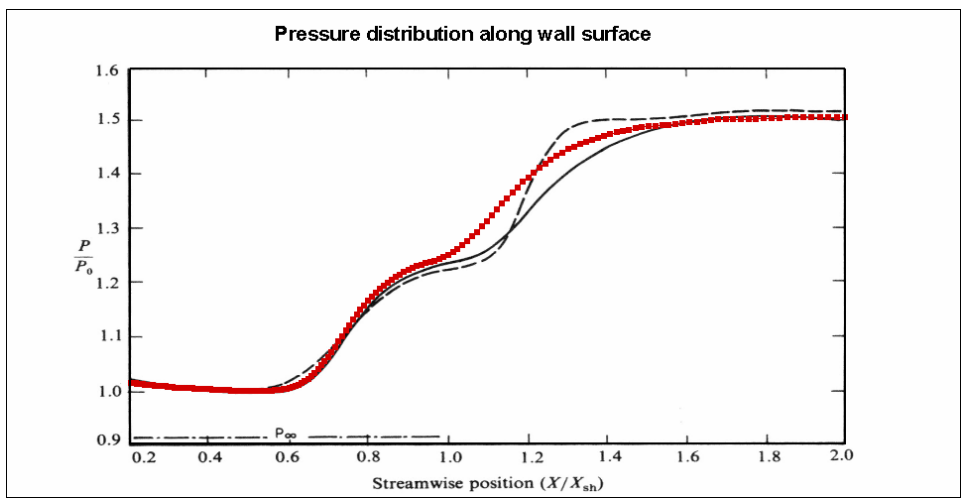

Figure 17 Comparison of pressure distribution on the wall surface. (The red one is our computation, the black dash one and solid one are Degrez's computation and experiment respectively)

Figures 15-16 show the distribution of pressure and density obtained by our computation. Figure 17 shows our numerical results agree well with the numerical results and are close to the experimental results given by Degrez et al (1987). Degrez et al favor their computational results addressed in their JFM paper. Figure 18 shows MWCS scheme has higher resolution for 2-D incident shock-boundary layer interaction. These results show our scheme can be extended to 2-D and 3-D problems.

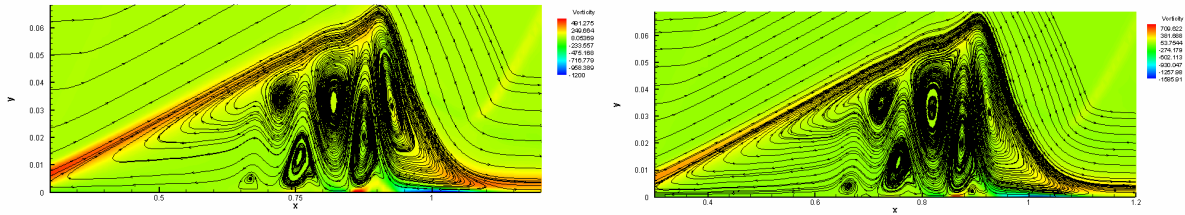

Figure 18 Verticity of 2-D incident shock-boundary layer interaction: (a) WENO (b) MWCS

## V. Conclusion

1) All three shock detectors, Harten's, WENO, and our new two step detector, can detect jump or corner points.
2) Harten's and WENO (Jameson's is similar) have difficulty distiguishing and misinterpreted high frequecy waves and critical points as shock, but our two step shock detector passed all checks for the above nine examples.
3) The new two step shock locator is very robust and has no need for manipulation of parameters.
4) The reason why our new detector is superior to others is that both Harten and WENO only consider first order derivative and second order derivative. The first order derivative is large for high frequency waves and the second order derivative is large for critical points. Both Harten's and WENO treat them as shock. However, Our new two step shock detector considers high order derivatives, fine (h) and coarse grids (2h), high order truncation error ratios and local left and right angle ratios.

Our two step shock detector never missed any shock/discontinuity including strong shock, weak shock, oblique shock and, moreover, never took high frequecy waves as shock to smear. Therefore, we believe we are in a very promising position to solve the most chalenging problem - to develop an effective shock/discontinuity locator. This topic is particularly important to shock-turbulence interaction, shock acoustic interaction, image process, porous media flow, multiple phase flow, etc. where both shock/discontinuity and small length scales are important.

## VI. Acknowledgement

This work was supported by US AFOSR Grant FA9550-05-1-0136 supervised by Dr. Fariba Fahroo and Air Force VA Summer Faculty Research Program supervised by Dr. Datta.Gaitonde.